\documentclass[a4paper,11pt]{article}
\usepackage{graphics}
\usepackage{jheppub}
\usepackage{epsfig}
\usepackage{color}
\definecolor{ourcolor}{rgb}{0.7, 0.25, 0.05}
\usepackage{float}
\usepackage{tabularx}
\newcolumntype{C}[1]{>{\centering\arraybackslash}p{#1}}
\expandafter\ifx\csname package@font\endcsname\relax\else
 \expandafter\expandafter
 \expandafter\usepackage
 \expandafter\expandafter
 \expandafter{\csname package@font\endcsname}%
 
\fi

\def\be{\begin{equation}}
\def\ee{\end{equation}}

\def\bea{\begin{eqnarray}}
\def\eea{\end{eqnarray}}
 
\title{\color{ourcolor}Explaining AMS-02 positron excess and muon anomalous magnetic moment in dark left-right gauge model}
\author{Tanushree Basak$^{1,}$}
\author{Subhendra Mohanty$^{1,}$}
\author{Gaurav Tomar$^{1,2,}$}

\affiliation{$^1$Physical Research Laboratory, Ahmedabad 380009, India }
\affiliation{$^2$Indian Institute of Technology, Gandhinagar 382424, India }

\emailAdd{tanu@prl.res.in}
\emailAdd{mohanty@prl.res.in}
\emailAdd{tomar@prl.res.in}

\abstract{
In a Dark left-right gauge model, the neutral component of 
right-handed lepton doublet is odd under generalized R-parity and thus the lightest one serves as the dark matter (DM) candidate. The coannihilation of the dark matter with the singly charged Higgs triplet  produces the correct relic abundance. We explain AMS-02 positron excess by the annihilation of 800 GeV dark matter into $\mu^+\mu^-\gamma$, through a t-channel exchange 
of the additional charged triplet Higgs boson. The DM is leptophilic which is useful for explaining the non-observation of any antiproton excess which would generically  be expected from DM annihilation. The large cross-section needed to explain AMS-02 also requires an astrophysical boost. In addition, we show that the muon $(g-2)$ receives required contribution from singly and doubly charged triplet Higgs 
 in the loops.}

 \keywords{Dark Matter, Dark left-right gauge model}
 
\begin{document}
\maketitle
\flushbottom

\section{Introduction}
\label{sec:intro}

There are two prominent experimental hints, which may point towards extension of Standard model (SM) for solution, 
namely the excess in positron flux observed by Alpha Magnetic Spectrometer (AMS-02)  \cite{Aguilar:2013qda, Accardo:2014lma} and the discrepancy between the measured 
\cite{Bennett:2006fi, Bennett:2008dy} muon $(g-2)$ and the SM 
prediction \cite{Aoyama:2012wk,Gnendiger:2013pva,Davier:2010nc,Hagiwara:2011af,Benayoun:2012wc,
Blum:2013xva,Miller:2012opa}. 
 The AMS-02 collaboration has released their recent data \cite{Accardo:2014lma} which shows that the positron fraction rises up to $\sim 425$ GeV. This is consistent with the PAMELA result about the positron fraction 
 \cite{Adriani:2008zr}. There is no corresponding excess in the antiproton flux over the cosmic ray background \cite{Adriani:2010rc,ams}. A leading explanation of the observed positron excess comes from the annihilation of dark matter particles into leptonic final states which results in a soft positron spectrum which can account for the AMS-02 data quite well. A population of nearby pulsars can provide an alternative explanation 
 \cite{Kamae:2010ad,Pato:2010im,Serpico:2011wg,Yin:2013vaa} for the positron excess reported by AMS-02, PAMELA. However in case of pulsar, an anisotropy is expected in the signal contributions as a function of energy due to the differing positions of the individual contributing pulsar, which falls nearly an order of magnitude below the current constraints from both AMS-02 and the Fermi-LAT \cite{Linden:2013mqa}.
 Another experimental signal is the discrepancy between the observed and the SM value of muon $(g-2)$. Beyond SM models where the leptons are preferentially coupled to the extra Higgs bosons can naturally address these two issues simultaneously. 

In this paper, we show that a variant of the left-right model called Dark left-right gauge model (DLRM) 
\cite{Khalil:2009nb} has the ingredients to explain these two experimental signals.
The alternative left-right symmetric model (ALRM) has been proposed in 1987 \cite{Ma:1986we,Babu}. One of its key advantages over the 
standard/conventional left-right model (LRM) \cite{Senjanovic:1975rk,Mohapatra:1979ia,Deshpande:1990ip,Pati:1974yy,Mohapatra:1974gc} is, it has no tree-level flavor changing neutral currents. Therefore, 
the $SU(2)_R$ breaking scale can be low and hence allows a possibility for $W_R^{\pm}, Z'$ gauge boson to be observable 
at collider experiments. Another variant of this ALRM is the dark left-right gauge model (DLRM) \cite{Khalil:2009nb,Aranda:2010sr}, 
which has both neutrinos and fermionic DM candidate. 
The neutral component of the right-handed lepton doublet `$n_R$' carries zero generalized lepton number ($\tilde L$) and is odd 
under the R-parity, $R=(-)^{3B+\tilde{L}+2j}$. Thus it can be made stable and a viable candidate for DM if it is the lightest R-odd 
particle in the spectrum. Additional Higgs triplet ($\Delta_R$) has been introduced to give mass to $n_R$. The annihilation of $n_R$ 
into muonic final states takes place through the t-channel  exchange of charged triplet Higgs ($\Delta_R^+$), which dominantly contributes to the relic abundance. The annihilation cross-section is maximized when the $\Delta^+_R$ and $\chi$ masses are close. 
But, in case of degeneracy in the masses of DM and $\Delta_R^+$, the relic abundance is described by co-annihilation. One of the motivations of this work is to explain the positron excess seen by AMS-02 \cite{Aguilar:2013qda, Accardo:2014lma} experiments 
through the annihilation of DM in the galactic halo. By choosing the $W_R^\pm, Z^\prime$ bosons heavier 
than the $\Delta_R$, we ensure that the DM is leptophilic which makes it ideal for explaining AMS-02 positron excess.
 But, the annihilation cross-section in this case is both helicity suppressed as well as p-wave suppressed at late times. To overcome this 
suppression, we have considered the mechanism of Internal bremsstrahlung (IB) \cite{Bergstrom:2008gr} in the 
DM annihilation process. Also we need an astrophysical boost $\sim \mathcal{O}(10^3)$, to get the required cross-section for AMS-02.
Another interesting aspect of this model is that the same Yukawa term $\Psi_R\Delta_R\Psi_R$, which produces muons from 
 DM annihilation also gives rise to the muon $(g-2)$ through singly and doubly charged triplet Higgs loop. 
We have shown that the same masses and coupling can be used to obtain both the relic abundance of DM and required 
$\Delta a_\mu =2.8 \times 10^{-9}$ within $1\sigma$ of the experimental value \cite{Bennett:2006fi, Bennett:2008dy}.
 Another model which can explain AMS-02 and muon $(g-2)$ has been constructed using a gauged horizontal symmetry \cite{Tomar:2014rya}.

The paper is organized as follows : In Section.\ref{sec:dlrm} we describe the details of the model; the dark matter part is discussed in 
Section.\ref{sec:dm}. The explanation of AMS-02 positron excess has been dealt in Section.\ref{sec:ams} and in Section.\ref{sec:g-2} the contribution to muon $(g-2)$ has been calculated in detail. Finally we summarize our result in Section.\ref{sec:conclu}.


\section{Dark Left-Right Gauge model}
\label{sec:dlrm}

\begin{table}[ht!]
\begin{center}
\begin{tabular}{|C{3cm}|C{6cm}|C{1cm}|C{1cm}|} \hline
Fermion  & $SU(3)_C \times SU(2)_L \times SU(2)_R \times U(1)$ & $S$ & $L$ \\\hline
      $\Psi_L=(\nu, e)_L$      &  (1,2,1,-1/2) & 1 & (1,1)\\\hline 
      $\Psi_R=(n, e)_R$      &  (1,1,2,-1/2) & 1/2  & (0,1)\\\hline 
      $Q_L=(u, d)_L$      &  (3,2,1,1/6) & 0 & (0,0)\\\hline 
      $Q_R=(u, x)_R$      &  (3,1,2,1/6) & 1/2 & (0,1) \\\hline 
      $d_R$      &  (3,1,1,-1/3) & 0 & 0\\\hline 
      $x_L$      &  (3,1,1,-1/3) & 1 & 1\\\hline 
\end{tabular}
\caption{Fermion content of DLRM model}
\label{tab}
\end{center}
\end{table}

We adopt the dark left-right gauge model (DLRM) \cite{Khalil:2009nb,Aranda:2010sr}, whose gauge group is given by, $SU(3)_C \times SU(2)_R \times SU(2)_L \times U(1)\times S$. Here an additional 
global $U(1)$ symmetry $S$ has been introduced such that after the spontaneous breaking of $SU(2)_R \times S$ the
generalized lepton number $\tilde{L}$ (defined as, $\tilde{L}=S-T_{3R}$) remains unbroken. The scalar sector of this model 
consists of a bi-doublet $\Phi$, two 
doublets ($\Phi_L, \Phi_R$) and two hypercharge `+1' triplets ($\Delta_L, \Delta_R$), denoted as,
\begin{center}
            $ \Phi=\begin{pmatrix}
                \phi_1^0 & \phi_2^+\\
                \phi_1^- &  \phi_2^0
               \end{pmatrix} $,
             $\Phi_{L,R}=\begin{pmatrix}
                   \phi_{L,R}^+\\
                    \phi_{L,R}^0
                  \end{pmatrix} $ and
            $ \Delta_{L,R}=\begin{pmatrix}
                \frac{\Delta_{L,R}^{+}}{\sqrt{2}} & \Delta_{L,R}^{++}\\
                \Delta_{L,R}^{0} &  -\frac{\Delta_{L,R}^{+}}{\sqrt{2}}
               \end{pmatrix} $
\end{center}
The fermionic sector (as shown in Table.\ref{tab}) consists of additional $SU(2)_R$ lepton ($\Psi_R$) and quark doublet ($Q_R$). Also it contains 
a quark singlet ($x_L$), which carries a generalized lepton number, `$\tilde{L}=1$'. 

\begin{figure}[t!]
\includegraphics[scale=0.6]{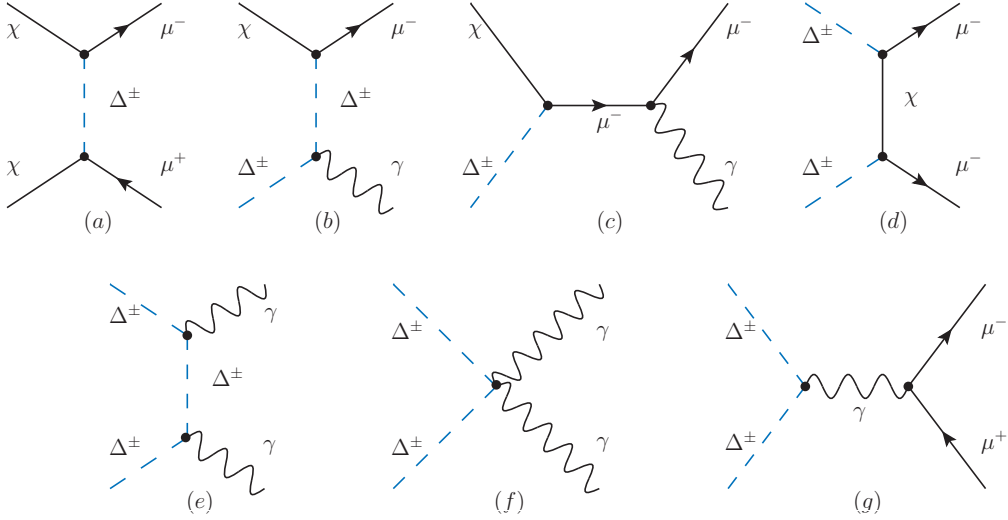}
\caption{\it Feynman diagrams of all dominant annihilation and coannihilation channels.}
\label{fig:dm_ann}
\end{figure}

The scalar potential contains all allowed (by $S$-symmetry) singlet combination like,
\begin{eqnarray}
 V &=& (m_1^2 \Phi^\dag \Phi + m_2^2 \Phi_L^\dag \Phi_L+m_3^2 \Phi_R^\dag \Phi_R+m_4^2 \Delta_L^\dag \Delta_L
 +m_5^2 \Delta_R^\dag \Delta_R) +\nonumber \\ 
  && \Phi_R^\dag \Delta_R \tilde{\Phi_R}+ \Phi_L^\dag \Phi \Phi_R+\textrm{Tr}(\tilde{\Phi}^\dag \Delta_L \Phi \Delta_R^\dag)+(quartic-terms)
\end{eqnarray}
From the minimization condition of the potential it is evident that there exists a solution with $\langle \phi_1^0 \rangle \equiv v_1=0$.
 The leptons and the up-quarks get mass through the Yukawa terms $\bar{\Psi_L}\Phi \Psi_R$ and $\bar{Q_L}\tilde{\Phi} Q_R$ respectively, when the neutral component of the bi-doublet gets vacuum expectation value (vev), i.e, $\langle \phi_2^0 \rangle=v_2$. 
Similarly the down quarks gets mass through the interaction $\bar{Q_L}\Phi_L d_R$. The triplet Higgses ($\Delta_{L,R}$) give masses to $\nu$ and $n$ respectively. Due to $S$-symmetry, terms like $\bar{\Psi_L}\tilde{\Phi} \Psi_R$ and $\bar{Q_L}{\Phi} Q_R$ are forbidden, which also ensures the absence of flavor-changing neutral currents at the tree-level. In addition, a generalized R-parity (defined as, $R=(-)^{3B+ \tilde{L}+2j}$) is imposed on this model, since $\tilde{L}$ is broken to $(-)^{\tilde{L}}$ when neutrinos acquire Majorana masses. This implies $n$, $x$, $W_{R}^\pm$, 
$\Phi_{R}^\pm$, $\Delta_{R}^\pm$ are odd under R-parity. One interesting feature of this model is that $W_{R}^\pm$-boson also carries generalized lepton number $\tilde{L}=\mp 1$, which forbids it from 
mixing with $W_{L}^\pm$-boson. This model also contains an extra $Z^\prime$-boson, but we have neglected the $Z-Z^\prime$ mixing, as the mass of the $Z^\prime$ is $\sim$ TeV and the mixing with $Z$ is small.


\section{Dark matter in DLRM}
\label{sec:dm}

By virtue of the $S$-symmetry the Yukawa-term $\bar{\Psi}_L\tilde{\Phi}\Psi_R$ is forbidden thus $n_R$ is not the Dirac mass 
partner of $\nu_L$. $n_R$ is termed as `scotino' \cite{Khalil:2009nb}, i.e, dark fermion and the lightest one is treated as a viable dark matter candidate. 
 The DM candidate is stable as an artifact of R-parity, under which it is odd. We choose $n_R^\mu \equiv \chi$, as the dark matter.  
The mass of DM is generated through the term $\Psi_R\Psi_R \Delta_R$. Here, we assume that $W_R^{\pm}, Z'$ 
gauge bosons are considerably heavier than $\Delta_R^+$.  
Therefore, the dominant annihilation channel of $\chi$ into leptonic final states (mainly $\mu^+\mu^-$) is through the t-channel exchange of $\Delta_R^\pm$ (as shown in Fig.\ref{fig:dm_ann}(a)). Since, the triplet Higgs does not couple with the quarks, the 
dark matter in this model is mostly leptophilic. Also there is no constraint on DM cross-section
 from direct detection experiments \cite{Aprile:2012nq, Akerib:2013tjd}.
   
   \begin{figure}[t!]
   \begin{center}
 \includegraphics[scale=1.0]{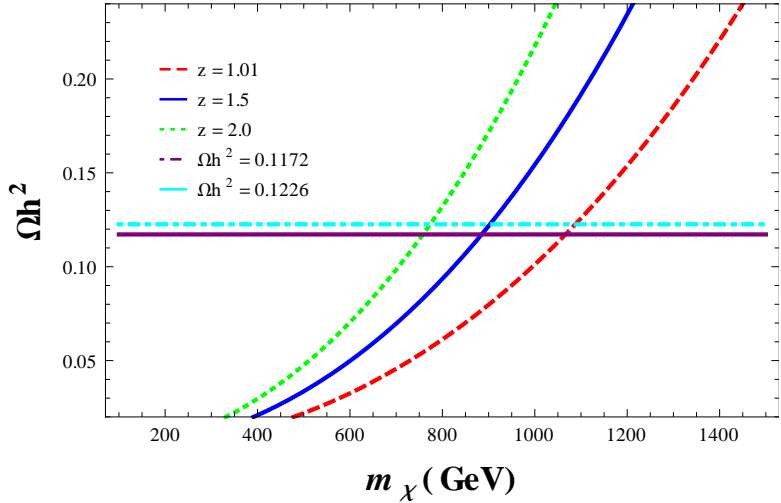}
 \caption{\it Plot of relic abundance as a function of DM mass, for $c_d=1.6$ and with different values of $z=1.01 \;
 (\textrm{red}), 1.5 \;(\textrm{blue}), 2.0\; (\textrm{green})$. The straight lines show the present value of 
 $\Omega h^2=0.1199 \pm 0.0027$ from Planck experiments \cite{Ade:2013zuv}.}
 \label{fig1}
 \end{center} 
\end{figure}
 

  Using partial-wave expansion, the annihilation cross-section can be written as, 
 $\langle\sigma v\rangle_{ann} \simeq a+6b/x_f$ 
where, a and b are the s-wave and p-wave contribution respectively. The s-wave part is helicity suppressed and is given by
\cite{Ellis:1998kh, Nihei:2002sc}, 
\begin{equation}
 a \simeq \frac{c_{d}^4}{32\pi m_\chi^2}\frac{m_f^2}{m_\chi^2}\frac{1}{(1+z)^2}
\end{equation}
whereas the p-wave contribution can be expressed as \cite{Cao:2009yy},
\begin{equation}
 b \simeq \frac{c_{d}^4}{48\pi m_\chi^2}\frac{(1+z^2)}{(1+z)^4}
\end{equation}
where, $c_d$ is the Yukawa-coupling between $\chi$, $\mu^-$ and $\Delta_R^+$. Again, 
 the ratio of RH-charged triplet mass to DM mass is denoted by, $z\equiv (m_{\Delta_R^+}/m_\chi)^2$. 
 Clearly, the s-wave contribution is negligible compared to the later part, which is velocity-suppressed today.
 
  If the masses of dark matter and the charged Higgs are nearly degenerate, i.e. $\delta m \sim T_f$ the 
 coannihilations \cite{Seckel,Edsjo,Ellis:1998kh} become important and relic density is no longer produced by thermal freeze-out. We have to take into account 
 cross-sections of processes like $\chi \Delta^+ \to \mu^+ \gamma$, $\Delta^+ \Delta^-\to \gamma \gamma$ and 
 $\Delta^+ \Delta^-\to \mu^+ \mu^- $ (as shown in Fig.\ref{fig:dm_ann}(b-f)). However, the contributions from the diagrams shown as Fig.1(d) and Fig.1(g) are less important since those are helicity-suppressed. 
 The effective cross-section is given by,
 \begin{equation}
 \sigma_{eff}v = \sum_{ij} \frac{n_i^{eq}n_j^{eq}}{(\sum_{k} n_k^{eq})^2}\sigma_{ij}v ,
 \end{equation}
where, $n_i^{eq}=g_i (\frac{m_i T}{2\pi})^{3/2} e^{-m_i/T}$.

 The analytic expression of the relic abundance can be formulated as \cite{Garny:2012eb,Garny:2013ama}
 \begin{equation}
 \Omega_{_{CDM}}h^2 \simeq \frac{\langle\sigma_{ann} v\rangle}{\langle\sigma_{eff} v\rangle}\bigg( \frac{T_{f0}}{T_f}\bigg) \bigg(\frac{m_\chi^2}{c_d^4}\bigg)\frac{(1+z)^4}{1+z^2}\textrm{GeV}^{-2}  
\end{equation}
where, $T_{f0}\simeq m_\chi/20$ is the temperature at the time of freeze-out and  $\langle\sigma_{ann} v\rangle$ is the 
annihilation cross-section without taking into account coannihilation.

\begin{figure}[t!]
   \begin{center}
 \includegraphics[scale=1.0]{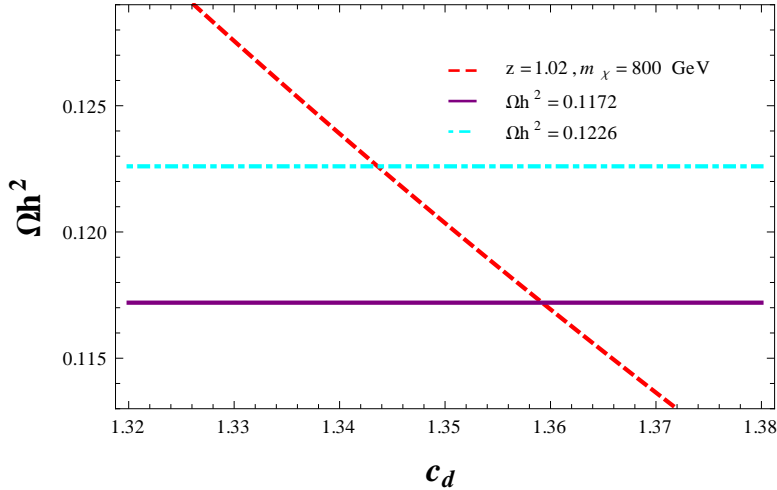}
 \caption{\it Plot of relic abundance as a function of coupling, for $m_\chi=800$ GeV and $z=1.02$. }
 \label{fig2}
  \end{center}
\end{figure}

 To produce the correct relic abundances, 
 one can tune the coupling $c_d$ and the ratio $z$. In Fig.\ref{fig1}, the 
relic abundance is plotted as a function of DM mass for $c_d=1.6$ but with different values of $z=1.01, 1.5, 2$. 
 The straight lines (solid and dashed) show the latest PLANCK data 
i.e, $\Omega_{_{CDM}} h^2 = 0.1199\pm 0.0027$ \cite{Ade:2013zuv}. 
We observe that as the ratio $z$ is increased, one requires lower values of dark matter mass in order to satisfy correct relic abundance.
We choose a specific set of benchmark point as, $m_\chi\sim 800$ GeV and $z=1.02$. We plot relic abundance, as shown in Fig.\ref{fig2},  for this particular choice of benchmark set. We obtain a narrow allowed range of coupling, i.e,  $1.343< c_d < 1.36$, which is consistent with relic abundance \cite{Ade:2013zuv}.

\subsection{Explanation of AMS-02 positron excess}
\label{sec:ams}

It has been shown that AMS-02 positron excess \cite{Aguilar:2013qda,Accardo:2014lma} can be explained by DM annihilation 
into $\mu^+\mu^-$ if the annihilation cross-section is $\sigma v \sim 10^{-24}\textrm{cm}^3\textrm{sec}^{-1}$ 
\cite{Cirelli:2008pk, DeSimone:2013fia}  for TeV scale DM. Such large cross-section needed to explain AMS-02 
through DM annihilation into `radiation' is constrained by recent Planck results \cite{planck}.  
 Therefore, the AMS-02 explanation necessarily requires an astrophysical boost \cite{Cumberbatch:2006tq,Lavalle:2006vb}. 
In DLRM, we have Majorana fermionic DM which implies that annihilation into fermionic final states is helicity
suppressed by a factor of $m_f^2/m_\chi^2$. 
As discussed earlier, the p-wave part of the annihilation cross-section is suppressed by the velocity squared
of the galactic DM particles today, which is typically $v_{today}\sim 10^{-3}$.  
One of the possibilities to evade the suppression is to make use of the Internal bremsstrahlung (IB) mechanism, where the emission  
of associated vector boson lifts the helicity suppression in the s-wave contribution to the annihilation cross-section
\cite{Bergstrom:2008gr,Bringmann:2012vr}. The process of IB incorporates both virtual internal bremsstrahlung (VIB) and the 
photons from final-state radiation (FSR). 
Therefore, we consider the annihilation of DM into $\chi\chi \to \mu^+\mu^-\gamma$ in the late universe (i.e. today), for which the  cross-section is given by \cite{Bergstrom:2008gr,Bringmann:2012vr},
 \begin{eqnarray}
 \langle\sigma v\rangle_{\mu^+\mu^-\gamma} & \simeq & \frac{\alpha_{em}c_{d}^4}{64\pi^2 m_\chi^2}\bigg\{(1+z)\bigg[\frac{\pi^2}{6}
 -\textrm{ln}^2\left(\frac{z+1}{2z}\right)-2\textrm{Li}_2\left(\frac{z+1}{2z}\right)\bigg] \\ 
  \nonumber && +\frac{4z+3}{z+1} +\frac{4z^2-3z-1}{2z} \textrm{ln}\left(\frac{z-1}{z+1}\right) \bigg\}
\end{eqnarray}
where, $\alpha_{em}$ is the fine-structure constant  and $\mbox{Li}_2(x)=\sum^{\infty}_{k=1}x^k/k^2$.

   \begin{figure}[t!]
   \begin{center}
 \includegraphics[scale=1.2]{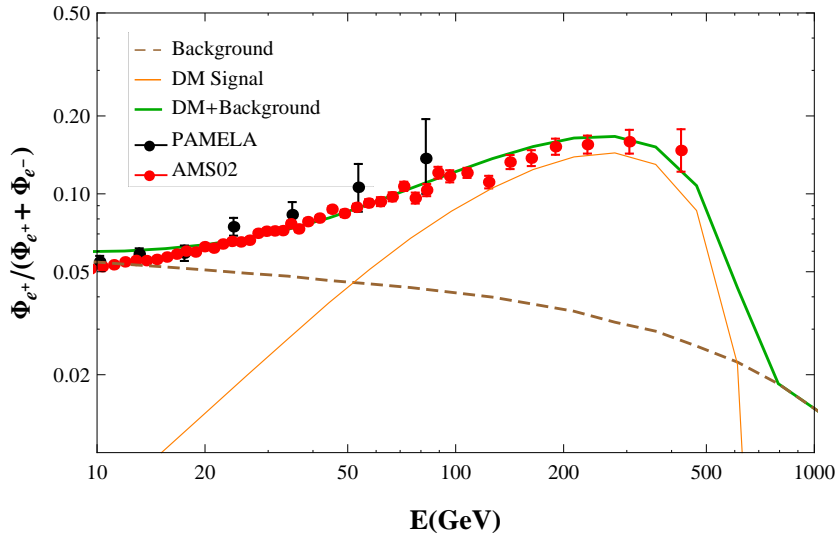}
 \caption{\it Prediction of the cosmic-ray positron fraction from dark matter annihilation into $\mu^+\mu^-$ final state. 
 The positron fraction spectrum is compared with the data from AMS-02 \cite{Aguilar:2013qda,Accardo:2014lma} and 
 PAMELA \cite{Adriani:2008zr}.}
 \label{fig:ams}
 \end{center}
\end{figure}

For generating the positron spectrum, $dN_e^+/dE$ from muon decay 
 $(m_\chi\sim~\mbox{800 GeV})$, we use the publicly available code PPPC4DMID \cite{8,9} and then we use GALPROP code 
 \cite{10,11} for the propagation of charged particles in the galaxy. The differential rate of production of primary positron flux per unit energy per unit 
 volume is  given by,
\begin{equation}
 Q_{e^+}(E, \vec r) = \frac{\rho^2}{2m^2_{\chi}}\langle \sigma v\rangle_{\mu^+\mu^-\gamma}\frac{dN_{e^+}}{dE}
 \label{st}
\end{equation}
where $\langle \sigma v\rangle_{\mu^+\mu^-\gamma}$ is the annihilation cross-section and $\rho$ denotes the density of dark matter
particle in the Milky Way halo, which we assume to be described by NFW profile \cite{12}. In GALPROP code \cite{10,11},
we set $D_0=3.6\times 10^{28}~\rm cm^2 s^{-1}$, $z_h = 4$ kpc and $r_{max}=20$ kpc, which are the diffusion coefficient,  
the half-width and maximum size of 2D galactic model respectively. We choose the nucleus injection index breaking at 9 GeV and
the values above and below its breaking are 2.36 and 1.82 respectively. Similarly in the case of electron, we choose injection
index breaking at 4 GeV and its spectral index above and below are 5.0 and 2.44 respectively with normalization flux 
$1.25 \times 10^{-8} \rm cm^{-2} s^{-1} sr^{-1} GeV^{-1}$ at 100 GeV.  Taking into account the chosen parameters, 
GALPROP \cite{10,11} solves the propagation equation, 
and we find the propagated positron flux.

 \begin{figure}[t!]
   \begin{center}
 \includegraphics[scale=1.2]{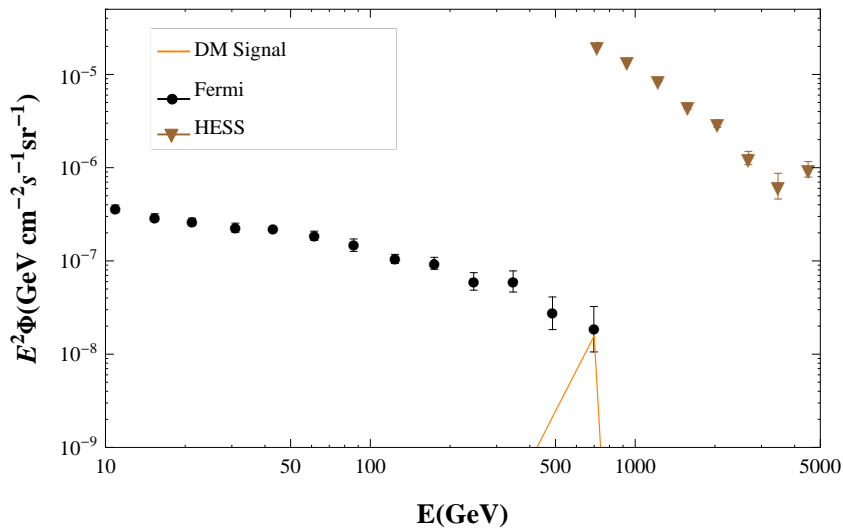}
 \caption{\it Predicted $\gamma$-ray spectrum is compared with Fermi LAT data \cite{Ackermann:2015tah}.  HESS measurement \cite{Aharonian:2008aa, Aharonian:2009ah} of $(e^+ + e^-)$ acts as upper bound on $\gamma$-ray 
flux in the 0.7-4 TeV range \cite{Kistler:2009xf}.}
 \label{fig:fermi}
  \end{center}
\end{figure}
In order to fit AMS-02 data \cite{Aguilar:2013qda, Accardo:2014lma}, the required annihilation cross-section 
in GALPROP code \cite{10,11} is
$\langle\sigma v\rangle_{\mu^+\mu^-\gamma}=8.8\times 10^{-25}\rm cm^3 s^{-1}$. But the internal bremsstrahlung process 
$(\chi\chi\rightarrow\mu^+\mu^-\gamma)$ gives the annihilation cross-section 
$\langle\sigma v\rangle_{\mu^+\mu^-\gamma}=1.37\times 10^{-28}\rm cm^3 s^{-1}$, using the benchmark set $m_\chi\sim 800$ GeV, 
$m_{\Delta_R^\pm}\sim 808$ GeV and $c_d\sim1.36$. 
It has been proposed in Ref.\cite{Cumberbatch:2006tq,Lavalle:2006vb} that local clumping at scales of 
$\sim 20$ kpc can enhance the positron flux (which arise from distances $< 20$ kpc) without changing the $\gamma$-ray or anti-proton flux \cite{Blanchet:2012vq} significantly. 
As an example of a astrophysical boost
we follow Raidal \cite{Hektor:2013yga} et al. and consider the positron and gamma ray flux from a local over-density. The usual number for the dark matter density in the solar neighbourhood is $\rho_0=0.4~\mbox{GeV}\mbox{cm}^{-3}$. If we assume a local over-density around the Sun by a factor $\sim 32$ over a size $0.4$ kpc then this forms a small fraction of the total volume and well within bounds of dark matter from solar system observations which allow $\rho \sim 15000 \rho_0$ \cite{Pitjev:2013sfa}. In such a case the flux of positron is \cite{Hektor:2013yga},
\begin{equation}
 \Phi_{e^+} = \frac{3\Gamma}{32\pi^2 K(E)R}\frac{dN_e^+}{dE},
\end{equation}
where $K(E)$ is Larmor radius and given as $K(E)=K_0(E/\mbox{GeV})^\delta$ with $\delta=0.85-0.46$ and $K_0=(0.016-0.0765)\mbox{kpc}^2/$Myr. The total dark matter annihilation rate $\Gamma$ is given by
\begin{equation}
 \Gamma = \frac{4\pi R^3}{3}\langle \sigma v\rangle \frac{1}{2}\frac{\rho^2}{M^2}.
\end{equation}
We get from AMS-02 positron data $\Phi_{e^+}=3.34\times 10^{-11}~\mbox{GeV}^{-1}\mbox{cm}^{-2}\mbox{sec}^{-1}\mbox{sr}^{-1}$ for $M\sim 400$ GeV, and with $R=370$ pc, $\delta=0.46$, $K_0=0.016~\mbox{kpc}^2/$Myr, $\langle \sigma v \rangle = 1.37 \times 10^{-28}~\mbox{cm}^3\mbox{sec}^{-1}$, and $dN_{e^+}/dE=0.003539~\mbox{GeV}^{-1}$, we get 
\begin{equation}
 \rho = 32~\rho_0.
\end{equation}
The effective astrophysical boost factor $(\sim \rho^2/\rho^2_0)$ is $\sim 6400$ which 
is required to explain the AMS-02 positron excess. The $\gamma$-ray photons (which mainly come from a distance larger than $0.4$ kpc) is smaller than the positron flux by a factor \cite{Hektor:2013yga}, 
\begin{equation}
 \frac{\Phi_\gamma}{\Phi_{e^+}} = \frac{2K}{R} \frac{dN_\gamma/dE}{dN_{e^+}/dE}.
 \label{eq:fcom}
\end{equation}
We conclude that a local over-density in dark matter by a factor 32 over a scale of 0.4 kpc can explain the positron excess observed in AMS-02 while at the same time not produce gamma rays in excess of what is observed by Fermi-LAT. 
In Fig.\ref{fig:ams}, we plot
the positron flux obtained from the GALPROP and compare it with observed AMS-02  \cite{Aguilar:2013qda, Accardo:2014lma} 
and PAMELA data \cite{Adriani:2008zr}. From Fig.\ref{fig:ams}, we observe that positron flux predicted from our model fits the data well.


 Since we are considering the internal bremsstrahlung process to lift the helicity suppression in the dark matter annihilation
cross-section, there will be primary photons in the final state  as well as secondary photons from muons. We also check the 
consistency of the predicted photon spectrum from this model with the observed data \cite{Ackermann:2015tah}. We have generated the $\gamma$-ray spectrum, i.e,  $dN_{\gamma}/dE$
using micrOMEGAs 3.3.9 code \cite{Belanger:2013oya}. We compare 
the output $\gamma$
spectrum with observed Fermi-LAT data \cite{Ackermann:2015tah}, which is shown in Fig.(\ref{fig:fermi}). 
The required cross-section for fitting AMS-02 positron excess, obtained in this model is consistent with the latest Fermi-LAT 4-year measurement of the gamma-ray background (see Fig.8 of Ref.\cite{Ackermann:2015tah}).
In Fig.\ref{fig:fermi}, we
have also shown the HESS measurement \cite{Aharonian:2008aa, Aharonian:2009ah} of $(e^+ + e^-)$, which acts as an upper bound
on $\gamma$-ray flux \cite{Kistler:2009xf} and clearly the $\gamma$-ray spectrum of our model is well below the upper limits.
 In this model, the dark matter does not annihilate into hadronic final states. Hence, there is no predicted excess of antiprotons,  which makes it consistent with the PAMELA \cite{Adriani:2010rc} and AMS-02 data \cite{ams}.


\section{Muon magnetic moment}
\label{sec:g-2}
\begin{figure}[t!]
 \begin{center}
 \includegraphics[scale=0.55]{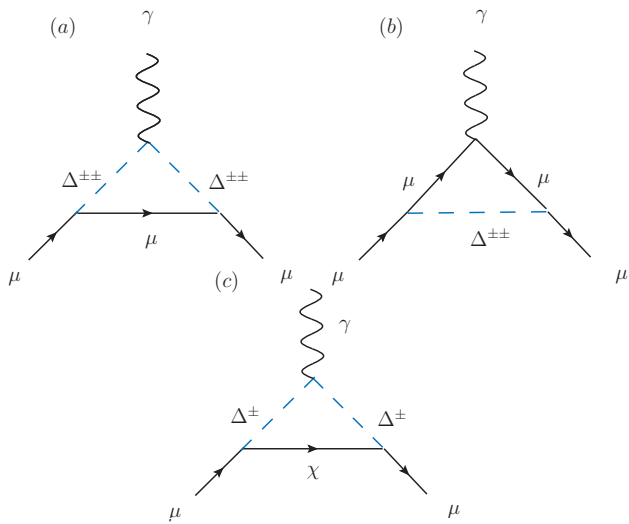}
\caption{\it Dominant Feynman diagrams of singly (c) and doubly (a,b) charged triplet scalar loops contributing to muon $(g-2)$.}
\label{fig:mm}
 \end{center}
\end{figure}
The muon magnetic moment is calculated by the magnetic moment operator, which is given as
 \begin{equation}
 {\cal L}_{MDM} = \frac{e}{2 m_{\mu}} F_2(q^2) \bar \psi_\mu \sigma_{\mu \nu} \psi_\mu F^{\mu \nu} 
 \end{equation}
where $m_\mu$ is the mass of the muon and $F_2(q^2)$ is the magnetic form factor. Here 
$\sigma_{\mu \nu} = \frac{i}{2} [\gamma_\mu,\gamma_\nu]$ and $F^{\mu \nu}$ is the field strength
of the photon field. 
The anomalous magnetic moment is related to $F_2$ as $\Delta a_\mu = F_2(0)$ for on-shell muon.

 In DLRM \cite{Khalil:2009nb}, there exist diagrams containing additional gauge bosons and charged triplet scalars which 
 give contributions to the muon magnetic moment. In the conventional left-right symmetric model 
 \cite{Pati:1974yy, Mohapatra:1974gc, Senjanovic:1975rk} with $g_L=g_R$, there are stringent 
 bounds from LHC on the masses of $SU(2)_R$ gauge bosons ($W^\pm_R, Z^\prime$), 
such that  $M_{W^\pm_R} \sim 2.5~\mbox{TeV},~M_{Z^\prime} \sim 3~\mbox{TeV}$ \cite{pdg2014}. Under these assumptions, the 
 contributions of heavy gauge bosons to muon $(g-2)$, have been neglected in comparison to the charged scalars. 
 Therefore, the interaction terms relevant to muon $(g-2)$ are $\psi_R\psi_R\Delta_R$ and $\psi_L\psi_L\Delta_L$. But, in the later term as the vev of $\Delta_L$ gives rise to neutrino masses, the Yukawa couplings are constrained to be sufficiently  small. Whereas, the former interaction term has no such restriction on the Yukawa coupling. Thus, we only consider the contribution from 
 $\Delta_R^+, \Delta_R^{++}$ loops to the anomalous magnetic moment of muon, as shown in Fig.(\ref{fig:mm}).

 The contribution from the doubly charged triplet Higgs (as shown in Fig. \ref{fig:mm}(a)-\ref{fig:mm}(b)) is given 
 by \cite{Leveille:1977rc},
 \begin{align}\label{chc}\nonumber
 [\Delta a_\mu]_{\Delta^{\pm\pm}} &= 4\times \bigg[\frac{2 m^2_\mu}{8\pi^2} \int^1_0 dx \frac{f^2_{\mu s}(x^3-x)
 + f^2_{\mu p}(x^3-2x^2+x)}{m^2_\mu ~(x^2-2x+1) + m^2_{\Delta^{\pm\pm}}x}\\
 & - \frac{m^2_\mu}{8\pi^2} \int^1_0 dx \frac{f^2_{\mu s}(2x^2-x^3) - f^2_{\mu p} x^3}
 {m^2_\mu ~x^2 + m^2_{\Delta^{\pm\pm}}(1-x)}\bigg]
 \end{align}
where $f_{\mu s}$ and $f_{\mu p}$ are the scalar and pseudoscalar couplings of charged triplet Higgs with the muon respectively. The factor of four in eq.(\ref{chc})
is a symmetry factor coming from the presence of two identical field in the interaction term 
($\psi_R\psi_R\Delta_R$). 
Similarly, the contribution from singly charged triplet Higgs $(\Delta^\pm_R)$, which is shown in diagram \ref{fig:mm}.(c), 
 given as,
 \begin{equation}\label{chc1}
 [\Delta a_\mu]_{\Delta^{\pm}}= \frac{m^2_\mu}{8\pi^2} \int^1_0 dx \frac{f^2_{\mu s}(x^3-x^2 +\frac{m_\chi}{m_\mu}
 (x^2-x))+f^2_{\mu p}(x^3-x^2 -\frac{m_\chi}{m_\mu}
 (x^2-x))}{m^2_\mu x^2 +(m^2_{\Delta^\pm}-m^2_\mu)x + m^2_\chi (1-x)}
 \end{equation}
 
 \begin{table}[ht!]
\begin{center}
\begin{tabular}{|C{3cm}|C{3cm}|C{3cm}|C{3cm}|C{3cm}|} \hline
$m_\chi$  & $m_{\Delta^\pm}$ & $m_{\Delta^{\pm\pm}} $ & $f_{\mu s}\simeq f_{\mu p}\equiv c_d$\\\hline
      800 GeV      &  808 GeV &  850 GeV & 1.36 \\\hline 
\end{tabular}
\caption{Value of parameters}
\label{tab2}
\end{center}
\end{table}

The choice of relevant parameters, in order to obtain the observed magnetic moment, has been depicted in  Table.\ref{tab2}. 
Here, we would like to mention that the same set of parameters is also required to explain the positron excess observed by AMS-02 experiment \cite{Aguilar:2013qda, Accardo:2014lma} and relic abundance of dark matter. After adding the contributions from eq.(\ref{chc}) and eq.(\ref{chc1}),  
we obtain
\begin{equation}
 \Delta a_\mu = 2.9 \times 10^{-9}
\end{equation}
which is in agreement with the experimental result \cite{Bennett:2006fi, Bennett:2008dy} within $1\sigma$.


\section{Discussion}
\label{sec:discu}

In dark left-right model, for explaining the AMS-02 positron excess, an astrophysical boost factor of $\sim6400$ is required. The large boost factor of this order is quite constrained in cold dark matter models \cite{Brun:2009aj}. In  \cite{Brun:2009aj}, authors have studied the positron and $\gamma$ flux from local dark matter clumps. They find that a local DM clump at 1 kpc distance with DM mass $\sim 650$ GeV and luminosity $L=3.4\times 10^9~\mbox{M}^2_\odot~\mbox{pc}^{-3}$ can explain the PAMELA positron excess (which is consistent with the AMS-02 positron excess). The calculated $\gamma$-flux $\Phi_\gamma = 10^{-6}~\mbox{cm}^{-2}\mbox{s}^{-1}$ is an order of magnitude larger than Fermi-LAT observation, which is $\Phi_\gamma = 10^{-7}~\mbox{cm}^{-2}\mbox{s}^{-1}$. The $\gamma$-flux observation is highly directional dependent, whereas positron flux is isotropic. So the positrons from the `point sources' of DM clusters will be observed but $\gamma$-rays can be missed if the telescope is not directed at the 
source. In this way it is possible to reconcile both the signals, but still the probability of such a large astrophysical boost factor is low as estimated from numerical simulation \cite{Brun:2009aj}.
\\
DLRM contains a number of singly and doubly charged scalars. According to the parameter space considered in this model, the dominant decay channel for $\Delta^{\pm\pm}_R$ is into same sign di-leptons, which is an important signal from the LHC perspective. The decay $\Delta^{\pm\pm}_R\rightarrow l^\pm l^\pm$, is constrained by CMS (ATLAS) collaboration, which exclude $m_{\Delta^{\pm\pm}}$ below 445 GeV (409 GeV) and 457 GeV (398 GeV) for $e^\pm e^\pm$ and $\mu^\pm \mu^\pm$ channels respectively \cite{ATLAS:2012hi,cms}. The singly charged scalar ($\Delta^{\pm}_R$) mass below 600 GeV (assuming, $BR : \Delta^+\rightarrow \tau^+ \nu_\tau = 1$) is ruled out at 95\% confidence level \cite{Chatrchyan:2012vca,atlas}. We considered $m_{\Delta^{\pm\pm}} \sim 808~\mbox{GeV and},~m_{\Delta^{\pm\pm}} \sim 850$ GeV for our calculations, which is above the exclusion limits.\\
$Z^\prime$ decays into SM fermions; $Z^\prime\rightarrow \ell^+\ell^-~(\ell=e,\mu)$ have been reported by CMS (ATLAS) collaboration, which put $M_{Z^\prime}\textgreater2.6~\mbox{TeV}~(2.9~\mbox{TeV})$ \cite{Chatrchyan:2012oaa,Aad:2014cka}. In DLRM, $M_{Z^\prime}$
and $M_{W^\pm_R}$ are related as \cite{Khalil:2009nb},
\begin{equation}
 M^2_{W_R}~\textgreater~\frac{(1-2x)}{2(1-x)}~M^2_{Z^\prime} + \frac{x}{2(1-x)^2}~M^2_{W_L},
\end{equation}
which gives, $M_{W_R}\textgreater 1.5~~\mbox{TeV}~(1.7~\mbox{TeV})$. Therefore, the contributions of heavy gauge bosons to relic density and muon $(g-2)$ will be small in comparison to charged scalars.
By choosing the coupling of $\Delta^{++}$ to $e^-e^-$ to be much smaller than the coupling with $\mu^-\mu^-$, we can evade 
the precession constraints from LEP \cite{Achard:2003mv}. 
We have assumed no flavor mixing through
$\Delta_R$, otherwise it will give rise to large contribution to $\mu\rightarrow e\gamma$ and $\mu\rightarrow eee$ process, which
is not observed \cite{pdg2014}.

\begin{figure}[t!]
 \begin{center}
 \includegraphics[scale=0.9]{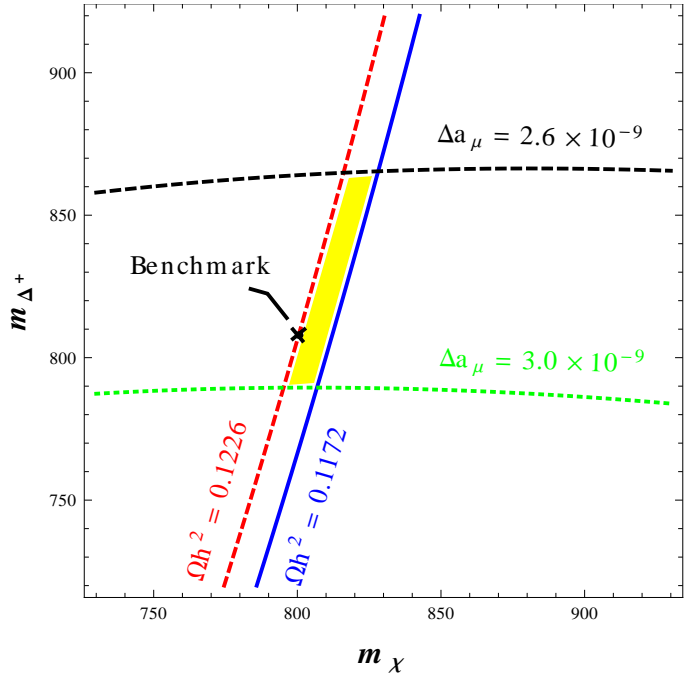}
\caption{\it Contours of $(g-2)$ and relic abundance in the plane of $m_\chi$ and $m_{\Delta^+}$ for $c_d=1.36$.}
\label{fig:param}
 \end{center}
\end{figure}

In this paper, we have claimed that we can accommodate both relic abundance and observed magnetic moment of muon in the same benchmark set. In Fig.\ref{fig:param}, we have shown the contours of muon $(g-2)$ and relic abundance in the plane of $m_\chi$ and $m_{\Delta^+}$ for $c_d=1.36$. The choice of particular coupling $c_d$ has been obtained from Fig.\ref{fig1}, where we have imposed constraints from relic abundance. The contours of correct relic abundance, consistent with PLANCK result \cite{planck}, shows the allowed range of masses of $m_\chi$ and $m_{\Delta^+}$. Now we observe that the contours of $g-2$ (within $1\sigma$) falls in the narrow band of parameter space favoured by relic density of dark matter. Therefore, we find that there exist a common parameter region (the shaded region in the middle as shown in Fig.\ref{fig:param}) of interest consistent with both relic abundance and anomalous magnetic moment. We have chosen a benchmark set, shown as a cross-mark, within the allowed region. However, the 
cross-section required to fit the AMS-02 positron excess narrows the parameter space as it requires $m_\chi \sim 800$ GeV.


\section{Conclusion}
\label{sec:conclu}

We have studied the DLRM model, in which the neutral component of RH-lepton doublet plays the role of a dark matter candidate. Thus, the Majorana fermionic DM candidate is stable as an artifact of generalized R-parity. In this model, we explain simultaneously the two experimental 
signatures $viz.$ AMS-02 positron excess and muon $(g-2)$ anomaly. The correct relic abundance 
of dark matter has been obtained through the coannihilation of the DM and the charged triplet Higgs. But the annihilation cross-section is helicity suppressed by a factor of $m_f^2/m_\chi^2$. Therefore we use the mechanism of internal bremsstrahlung to lift the helicity suppression. In order to explain AMS-02 positron excess, we have considered the annihilation of the dark matter into 
$\mu^+\mu^-\gamma$ with an additional local astrophysical boost factor  $\sim 6400$. Here we would like to mention that the constraints
from distant objects such as dwarf galaxies, and distant epochs such as
the cosmic microwave background have been evaded by virtue of having a small underlying cross section and relying on a large local boost factor. The prediction of positron excess of this model is in good agreement with PAMELA, AMS-02 data. We also obtain the required 
contribution to muon $(g-2)$ through the additional charged triplet loops and using the same set of parameters.
 We predict a downturn in the AMS-02 positron spectrum and a cut-off around 500 GeV. In addition as a signature of internal bremsstrahlung there is a peak in the gamma rays spectrum $\sim 0.8$ TeV which is consistent with both Fermi-LAT $\gamma$-ray observations and HESS upper bound.

 \section*{Acknowledgement}
 We would like to thank the anonymous referee for his/her constructive comments on the manuscript.


\end{document}